\input{aipcheck}

\documentclass[
    ,final           
  ]
  {aipproc}

\layoutstyle{8x11single}
\setcitestyle{numbers}

\begin{document}

\title{Dependence on the quark masses of elastic phase shifts and light resonances 
within standard and unitarized ChPT}


\author{J. Nebreda}{
}

\author{J.R. Pel\'aez}{
 address={Dept. F\'isica Te\'orica II. Universidad Complutense, 28040, Madrid. Spain}
 }

\begin{abstract}
We study the dependence of the $\pi \pi$ scattering phase shifts on the light quark mass in both standard and unitarized SU(2) Chiral Perturbation Theory (ChPT) to one and two loops. We then use unitarized SU(3) ChPT to study the elastic $f_0(600)$, $\kappa(800)$, $\rho(770)$ and $K^*(892)$ resonances. 
The quark masses are varied up to values of interest for lattice studies. 
We find a very soft dependence on the light quark mass of the $\pi \pi$ phase shifts at one loop and lightly stronger at two loops and a good agreement with lattice results. 
The SU(3) analysis shows that the properties of the $\rho(770)$ and
$K^*(892)$ depend smoothly on the quark mass whereas the
scalar resonances present a non-analyticity at high quark masses. 
We also confirm the lattice assumption of quark mass independence
of the vector two-meson coupling that, however,
is violated for scalars.  
\end{abstract}

\maketitle

\section{Introduction}
Although QCD is well established as the theory of strong interactions, 
the hadronic realm is beyond the reach of perturbative calculations.
In that regime, lattice methods are a useful tool to
calculate QCD observables, but there are still few results at small quark masses.
On the other hand, Chiral Perturbation Theory (ChPT) provides the quark mass dependence 
of the meson-meson scattering amplitudes at low energies. This allows us to increase the quark mass in the amplitudes in order to compare with lattice results. In this talk we first study the phase shift dependence on the averaged $u$ and $d$ quark mass, $\hat{m}$, in standard one and two-loop SU(2) ChPT\cite{Bijnens:1995yn}. This approach has the advantage of being completely model independent. Then we use the elastic Inverse Amplitude Method (IAM) to unitarize the amplitudes, which allows first, to calculate the phase shift quark mass dependence up to higher values of $\hat{m}$ and next, to generate poles on the second Riemann sheet and study their evolution. This technique was applied in~\cite{Hanhart:2008mx} to calculate the 
$\hat{m}$ dependence of the $f_0(600)$ and
$\rho(770)$ resonances. Here we extend this study to include the strange quark within a unitarized SU(3) ChPT formalism \cite{Nebreda:2010wv}, so that we can also generate the $K^*(892)$ and $\kappa(800)$ resonances and vary both the light and the strange quark masses.

\vspace*{-.4cm}
\section{Quark mass dependence in Standard ChPT} 

\begin{figure}
\begin{tabular}{c}
  \includegraphics[scale=1.3]{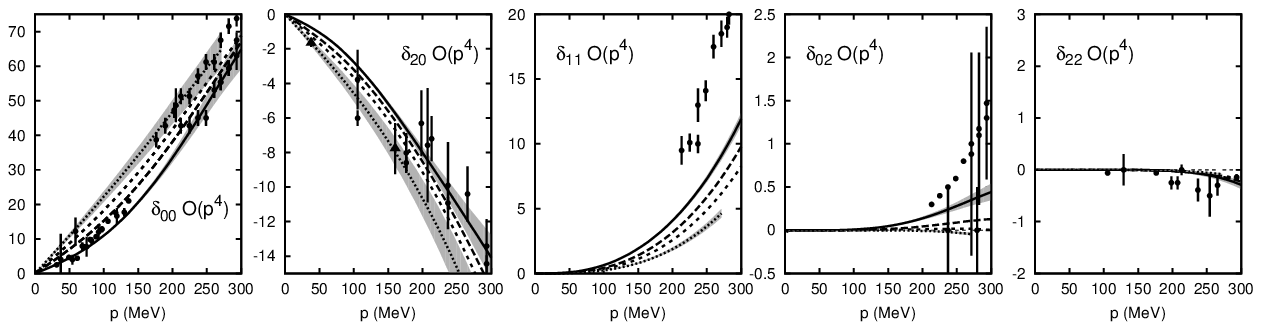}\\
  \includegraphics[scale=1.3]{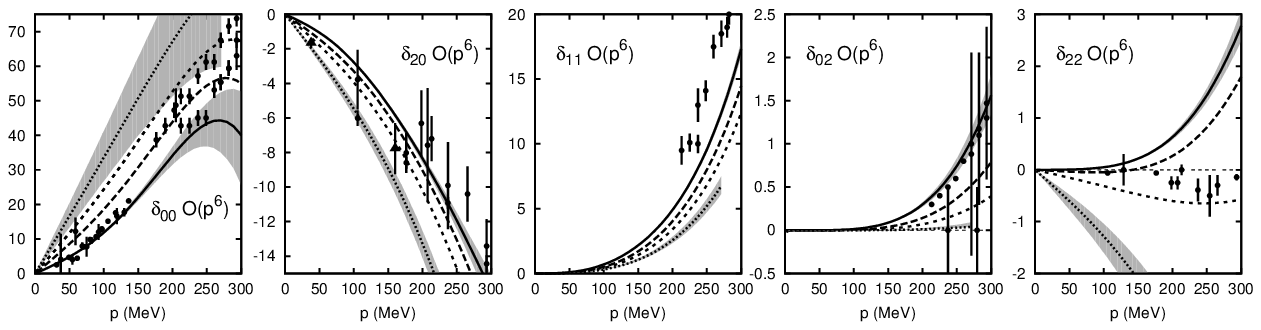}
  \caption{}
  \end{tabular}
  \label{fig:NUphaseshifts}
    \caption{$\pi \pi$ phase shifts from standard ChPT up to one loop (first row) and to two loops (second row). Different lines stand for different pion masses: continuous, long dashed, short dashed and dotted for $M_\pi=139.57,\, 230,\, 300$ and 420 MeV respectively. Because of applicability considerations, we do not plot phase shifts beyond $\sqrt{s}=1$ GeV. Since the lines are too close to each other, we only show error bands for the lightest and heaviest masses. Experimental data (circles) come from~\cite{experimentaldata}. Lattice results for the I=2, J=0 channel (triangles) come from~\cite{Sasaki:2008sv}.
    }
\end{figure}

We use the SU(2) scattering amplitudes in \cite{Bijnens:1995yn} and the LECs in \cite{Colangelo:2001df} to study the dependence of the $\pi \pi$  phase shifts on $\hat m$. In Fig.~\ref{fig:NUphaseshifts} we show preliminary results for phases on different channels for different pion masses, to one and two loops (first and second row respectively). They are plotted as a function of the center of mass momentum -and not of the energy- in order to subtract the shift of the thresholds. Since the applicability of plain ChPT is limited to low energies, we do not plot phase shifts beyond $\sqrt{s}=1$ GeV.  In order to compare with lattice results~\cite{Sasaki:2008sv}, we rise $M_\pi$ up to 420 MeV, although most likely the IAM is not reliable in that region. A Montecarlo gaussian sampling based on the errors of the LECs has been used to calculate the error bands.\\

We find that the dependence of the phase shifts on $\hat{m}$ is very soft at one loop and somehow stronger at two loops, specially for the I=2, J=2 channel, and that the results are compatible with those from lattice in the I=2, J=0 channel~\cite{Sasaki:2008sv}.

\vspace*{-.4cm}
\section{Quark mass dependence in Unitarized ChPT} 

We use now the IAM to unitarize our amplitudes. This procedure extends their applicability up to the resonance region and generates poles in the second Riemann sheet, which relate to resonances through the usual Breit-Wigner identification $\sqrt{s_{pole}}\equiv M-i\,\Gamma/2$. Furthermore, the coupling of the resonance to two mesons is given by the residue of the amplitude at the pole position. 
\vspace*{-.3cm}

\begin{figure}
\begin{tabular}{c}
  \includegraphics[scale=1.55]{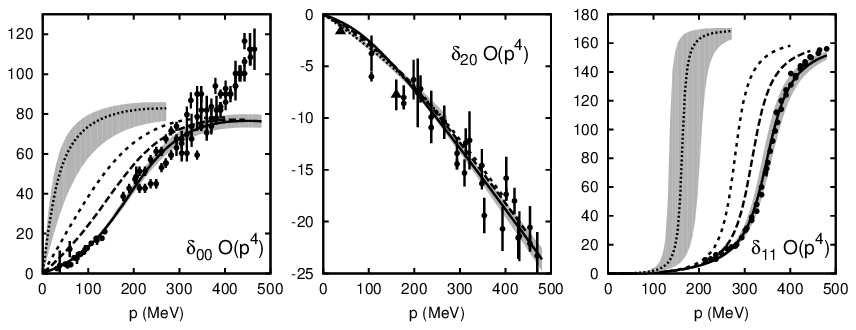}\\
  \includegraphics[scale=1.55]{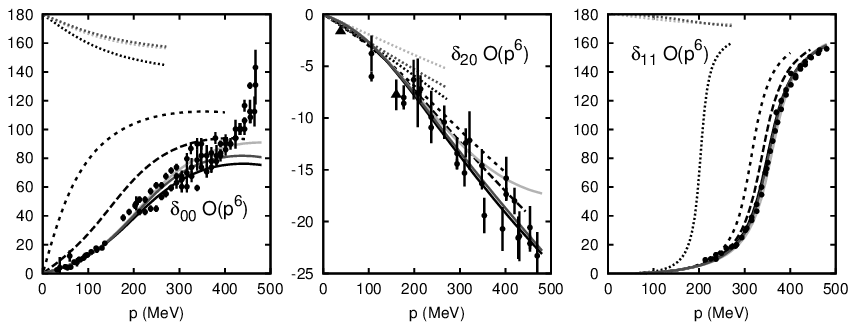}
  \caption{Uphaseshifts}
  \end{tabular}
  \label{fig:Uphaseshifts}
    \caption{$\pi \pi$ phase shifts from unitarized ChPT up to one loop (first row) and to two loops (second row). The conventions are as in Fig.~\ref{fig:Uphaseshifts} except that, in the case of the two loop analysis, instead of error bands we show results for three different sets of LECs\cite{Riospelaez:prep} (black, dark gray and light gray).}

\end{figure}

\paragraph{Phase shifts} In Fig.\ref{fig:Uphaseshifts} we show preliminary results for $\pi \pi$ phase shifts (note that the one and two-loop IAM cannot be used for D-waves) for different pion masses. For the one loop analysis (first row) we used the LECs in \cite{Hanhart:2008mx} and found that the dependence is again quite soft, specially for the I=2, J=0 channel. For the two loop analysis (second row) we used some preliminary sets of LECs~\cite{Riospelaez:prep}. Once more the dependence on $M_\pi$ is stronger at two loops than at one loop.

\begin{figure}
\begin{tabular}{@{\extracolsep{-10pt}}cc}
  \includegraphics[scale=0.65]{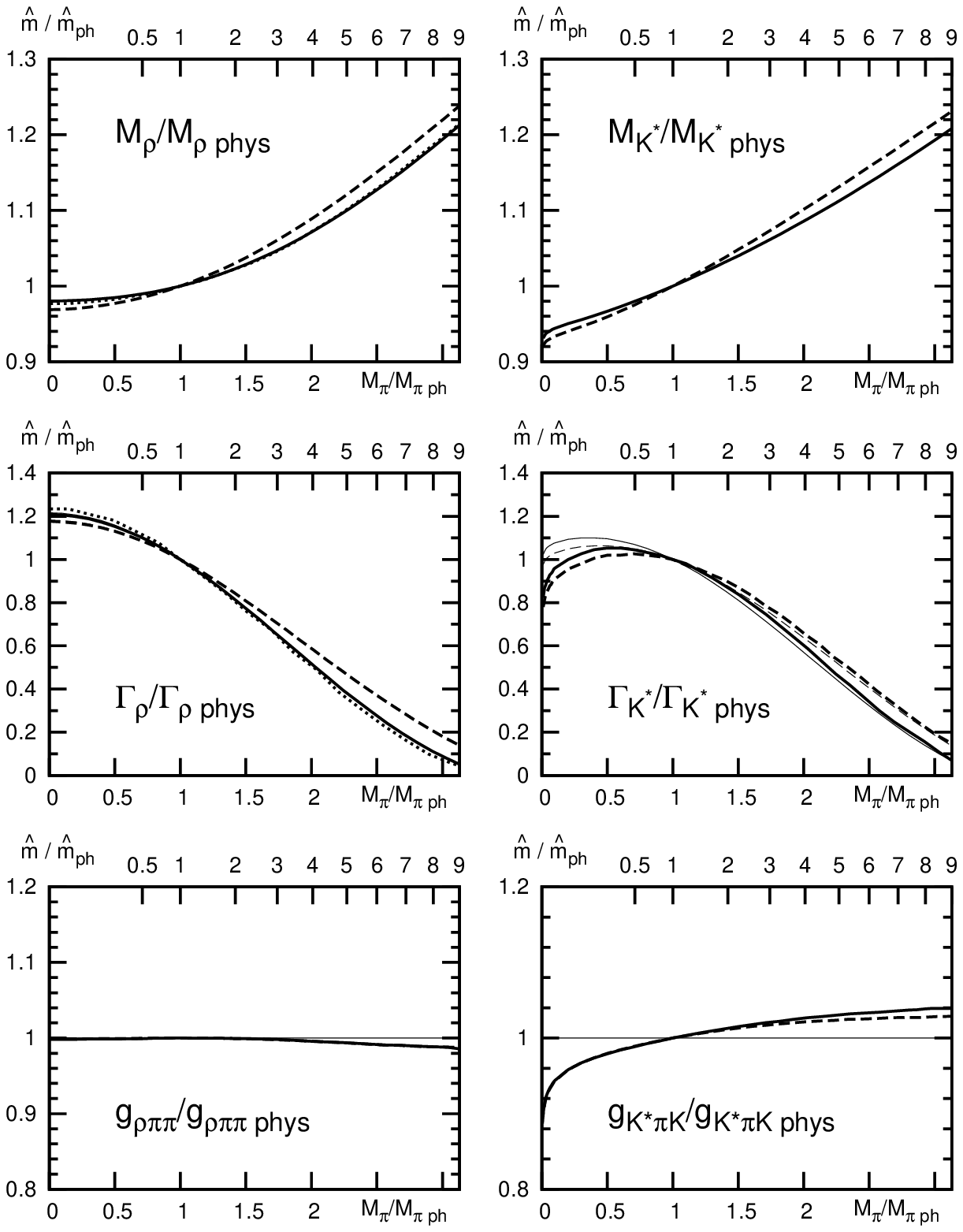} &   \includegraphics[scale=0.65]{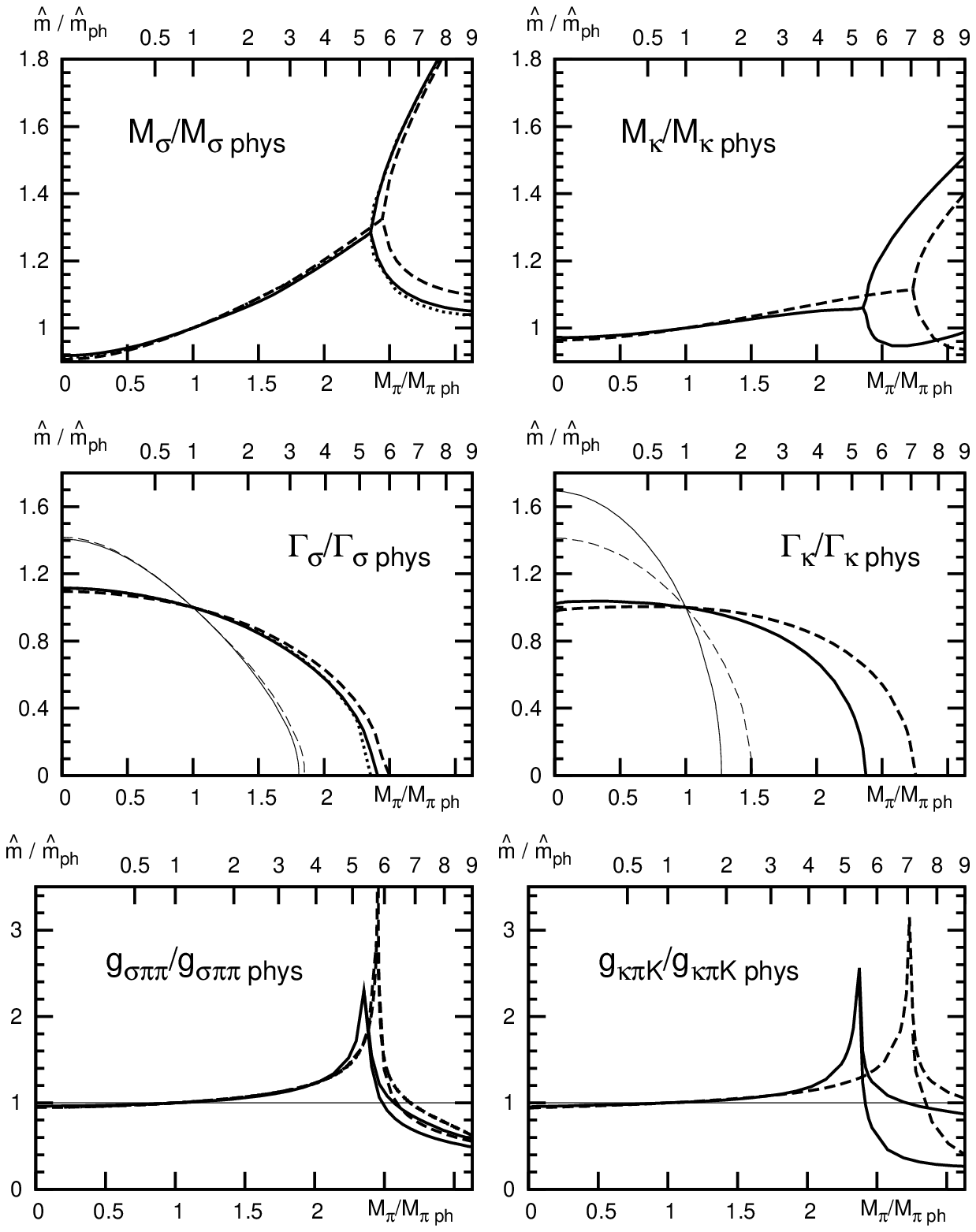}\\
  [5ex]
  \includegraphics[scale=0.65]{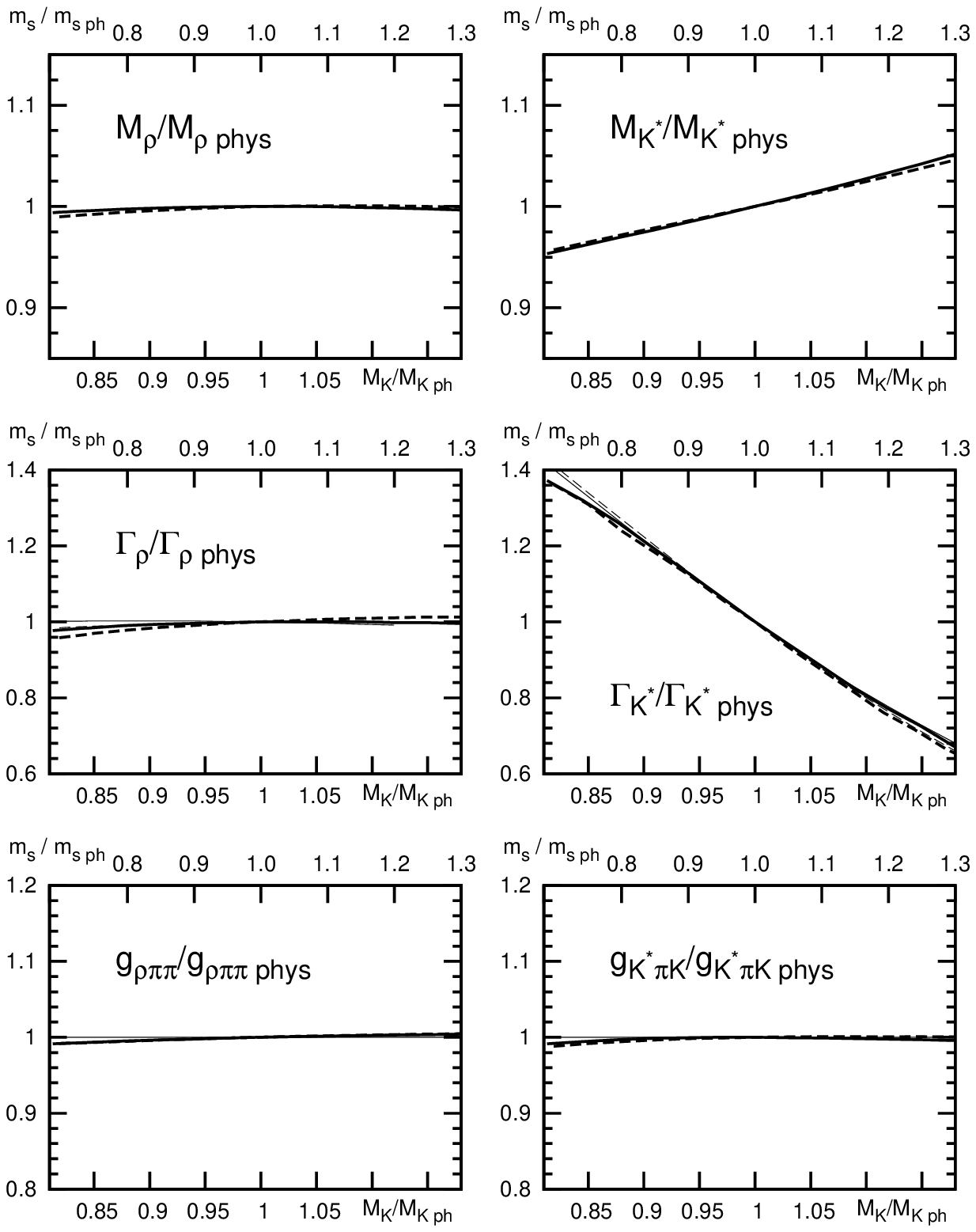} &   \includegraphics[scale=0.65]{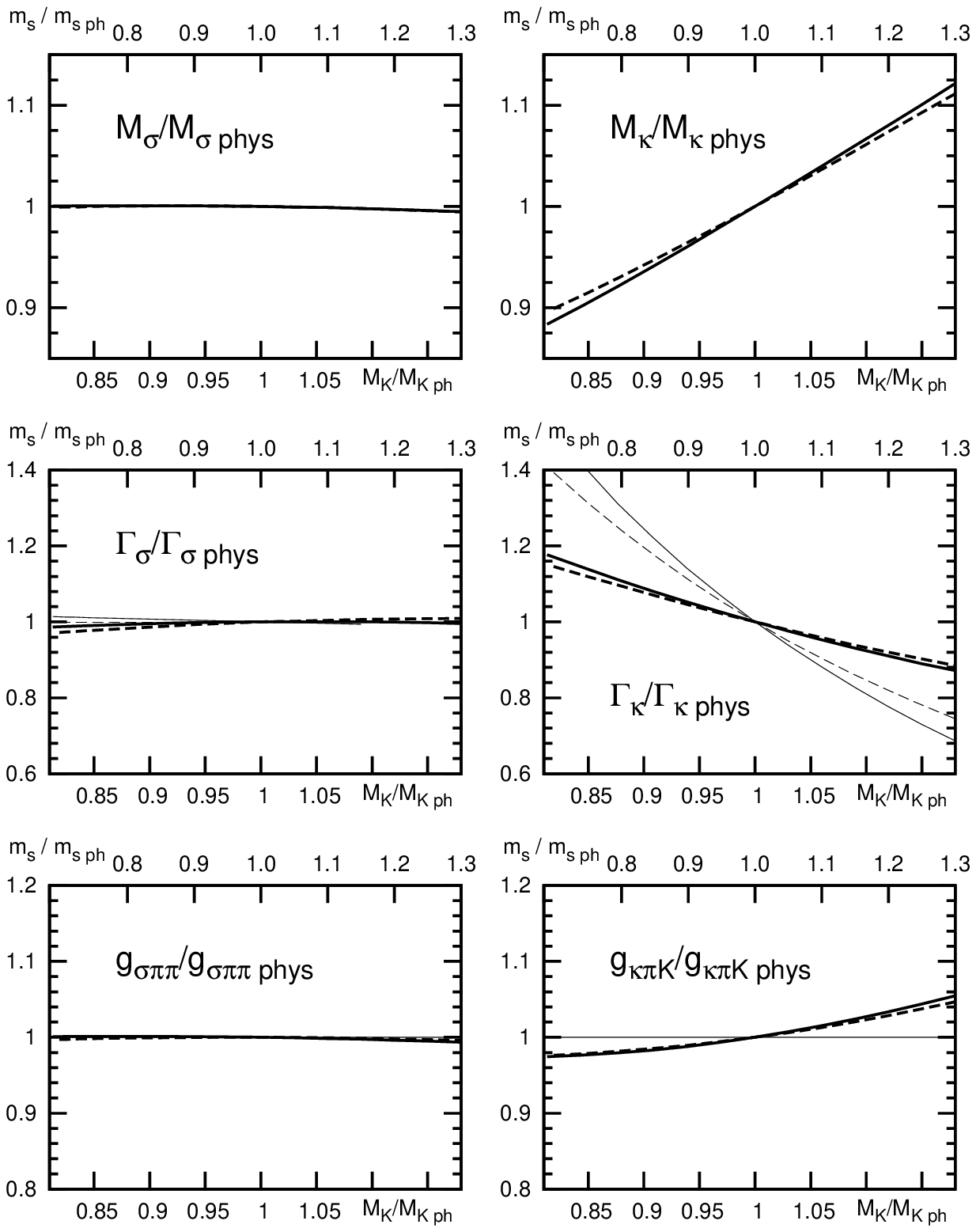}
  \caption{}
  \end{tabular}
  \label{fig:resonances}
    \caption{Dependence of the $\rho(770)$, $K^*(892)$, $f_0(600)$ 
 and $\kappa(800)$ mass, width and coupling to two mesons with respect to
$\hat{m}$ (horizontal upper scale) or $M_\pi$ (horizontal lower scale) for the three upper rows and with respect to $m_s$ (horizontal upper scale) or $M_K$ (horizontal lower scale) for the three lower rows.
Note that all quantities are normalized to their 
physical values. 
The thick continuous and dashed lines correspond to
Fit I and Fit II in Ref.~\cite{Nebreda:2010wv}, respectively. The
$\rho(770)$ and $f_0(600)$ dependence on the light quark mass is very 
compatible with that in \cite{Hanhart:2008mx} (dotted lines).
Continuous (dashed) thin lines show the dependence of
the widths from the change of phase space only, assuming a constant
coupling of the resonances to two mesons, using Fit I (II).
    }
\end{figure}

\vspace*{-.3cm}
\paragraph{Resonance dependence on the light quark mass \cite{Nebreda:2010wv}}
The vector resonances are well established $q \bar q$ states belonging to 
an SU(3) octet. In the three upper rows of Fig.\ref{fig:resonances} we show the dependence of 
the $\rho(770)$ (first column) and $K^*(892)$ (second column) on the light quark mass. The results obtained for the former are very consistent
with those in SU(2)~\cite{Hanhart:2008mx} (dotted line) and the estimations 
for the two first coefficients of the $M_\rho$ chiral expansion \cite{bruns}. Both resonances behave very similarly: their masses increase smoothly, but much slower than 
$M_\pi$. As a consequence there is a strong phase space suppression which accounts by itself 
for the width decrease, without a dynamical effect
through the couplings $g_{\rho\pi\pi}$ and
$g_{K^*\pi K}$, that are remarkably constant,
which is an assumption made in lattice studies
of the $\rho(770)$ width \cite{Aoki:2007rd}.\\
On the other hand, the $f_0(600)$, or sigma, and the $\kappa(800)$ scalar mesons 
are still somewhat controversial since their huge width makes their experimental identification complicated. The third and fourth columns show their dependence on $\hat{m}$. As before, the results for $f_0(600)$ are 
in good agreement with~\cite{Hanhart:2008mx}.
The most prominent feature of the scalars behavior
is the splitting of the mass into two branches. This happens when the associated pair of conjugated poles in the second Riemann sheet, which are approaching each other as the quark mass increases, join in a single pole below threshold to
split again and remain in the real axis.
Quantitatively, the growth of the $f_0(600)$ mass before the ``splitting point'' is 
much faster than that of the $\kappa(800)$.
Their width decrease cannot be attributed to the phase space reduction now, as show by the thin lines,
because their coupling to two mesons depend strongly on the quark mass. 

\paragraph{Resonance dependence on the strange quark mass~\cite{Nebreda:2010wv}} We will only vary the strange quark mass
in the limited range $0.8<m_s/m_{s\,\rm phys}<1.4$ to ensure that the kaon does not become
too heavy to spoil the ChPT convergence nor too light to require a coupled channel
formalism.
In the three lower rows of Fig.~\ref{fig:resonances} 
we show the kaon (or strange quark) mass dependence 
of the $\rho$,  $K^*(892)$, sigma and $\kappa(800)$.
As it could be expected, the properties of the $\rho(770)$ and sigma non-strange resonances
are almost independent of the strange quark mass within the range of study. 
Obviously, the $K^* (892)$ and the $\kappa(800)$ show a strong $m_s$ dependence.
As the kaon mass is made heavier, their masses grow much
faster than they did when increasing the light quark mass, but still much slower than the kaon mass.
As before, in the case of the vector resonance the width
decreases almost exactly as it would be expected from phase space suppression only and its coupling to $K\pi$ is almost constant, while the scalar width decrease deviates significantly from that behavior, in agreement with $g_{\kappa\pi  K}$ depending quite strongly on the strange 
quark mass.

\vspace*{-.3cm}
\section*{Acknowledgments}

\vspace*{-.1cm}
Work partially supported by Spanish MICINN: FPA2007-29115-E,
FPA2008-00592 and FIS2006-03438,
U.Complutense/ Banco Santander grant PR34/07-15875-BSCH and
UCM-BSCH GR58/08 910309 and the EU-Research Infrastructure
Integrating Activity
``Study of Strongly Interacting Matter''
(HadronPhysics2, Grant n 227431)
under the EU Seventh Framework Programme.

\end{document}